\documentclass{article}
\usepackage{spconf,amsmath,graphicx}
\usepackage{enumitem}
\setlist{topsep=0pt, leftmargin=*}

\usepackage{color,xspace,listings,comment}
\usepackage[colorinlistoftodos]{todonotes}

\title{QuApprox: A Framework for Benchmarking the Approximability of Variational Quantum Circuit\vspace{-5pt}}
\name{Jinyang Li$^{\dag,\pounds}$ \qquad Ang Li$^{\S}$ \qquad Weiwen Jiang$^{\dag,\pounds}$
  \vspace{-5pt}}
  \address{
  $^{\dag}$Department of Electrical and Computer Engineering, George Mason University, VA, USA. \\$^{\pounds}$ Quantum Science and Engineering Center, George Mason University, VA, USA. \\$^{\S}$Pacific Northwest National Laboratory, WA, USA. \\ \{jli56, wjiang8\}@gmu.edu 
      }
\begin{document}
%
\maketitle

\thispagestyle{empty}

\begin{tikzpicture}[remember picture,overlay]
   \node[anchor=south, yshift=10pt, text width=.8\paperwidth] at (current page.south) {
       \centering
       \footnotesize
       © 2024 IEEE. Personal use of this material is permitted. Permission from IEEE must be obtained for all other uses, in any current or future media, including reprinting/republishing this material for advertising or promotional purposes, creating new collective works, for resale or redistribution to servers or lists, or reuse of any copyrighted component of this work in other works.
   };
\end{tikzpicture}
\begin{abstract}
Most of the existing quantum neural network models, such as variational quantum circuits (VQCs), are limited in their ability to explore the non-linear relationships in input data. This gradually becomes the main obstacle for it to tackle realistic applications, such as natural language processing, medical image processing, and wireless communications. Recently, there have emerged research efforts that enable VQCs to perform non-linear operations. However, it is still unclear on the approximability of a given VQC (i.e., the order of non-linearity that can be handled by a specified design). In response to this issue, we developed an automated tool designed to benchmark the approximation of a given VQC. The proposed tool will generate a set of synthetic datasets with different orders of non-linearity and train the given VQC on these datasets to estimate their approximability. Our experiments benchmark VQCs with different designs, where we know their theoretic approximability. We then show that the proposed tool can precisely estimate the approximability, which is consistent with the theoretic value, indicating that the proposed tool can be used for benchmarking the approximability of a given quantum circuit for learning tasks.
\end{abstract}
\section{Introduction}
Quantum machine learning (QML) combines quantum computing with machine learning to improve algorithms. With its unique features such as superposition and entanglement, QML is able to perform highly parallel and efficient computations.
One popular approach is to use variational quantum circuits (VQCs) \cite{ezhov2000quantum,dunjko2018machine,schuld2018supervised,biamonte2017quantum}, which are a type of quantum model that can be trained using classical optimization techniques. VQCs can be used to classify data, and they have shown promising results on simple datasets such as MNIST. However, VQCs have limited capabilities when it comes to handling non-linear relationships between input and output data \cite{cerezo2022challenges}. This is because VQCs are more similar to a linear model as the computation circuit itself can be represented as a parameter matrix used to linearly transform the statevector to another. They struggle to effectively capture the complex non-linear relationships present in many real-world datasets.

Motivated by that, research on feeding nonlinearity into the pipeline of quantum learning has started drawing attention \cite{gili2023introducing,li2023novel,parthasarathy2021quantum,yan2020nonlinear}, while the effective validation of whether models can gain enhancement from the existing techniques needs to be further explored. In other words, regarding a given VQC design, its approximability(i.e. to what extent it can handle the nonlinear) is unclear. In order to measure their capability to handle non-linearity, in this paper, we propose an automated tool that is designed to benchmark the approximation of a given VQC. The dataset generator within the tool will create a couple of dataset groups where each group consists of multiple synthetic datasets with a certain non-linearity degree. In that way, we can train and test the given VQC on these dataset groups with different non-linearity degrees to obtain the corresponding approximability. Then, in our experiments, we take different VQC designs for benchmarking where their theoretical approximability is known. Our results demonstrate that our proposed tool accurately estimates the approximability, aligning with the theoretical values. This shows that our tool is effective for benchmarking the approximability of specific VQC in learning tasks.


\begin{figure*}[t]
\centering
\includegraphics[width=0.95\linewidth]{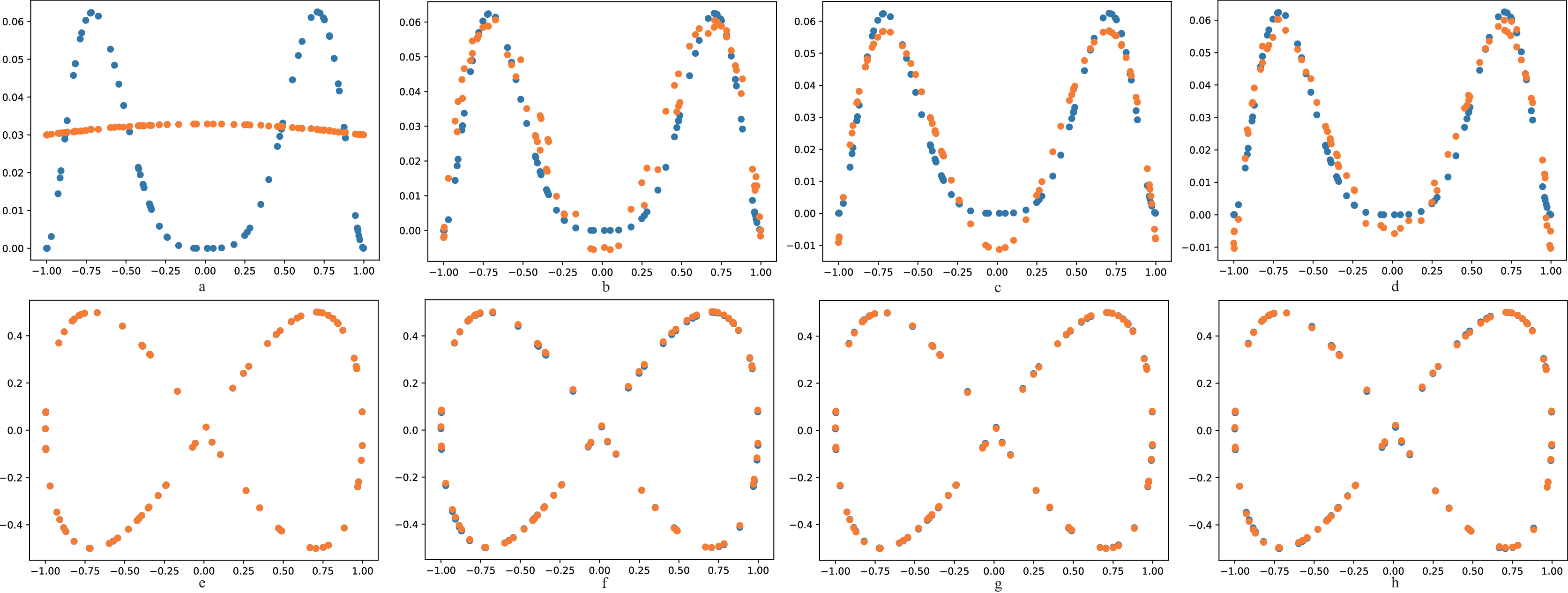}
\caption{From top-left to top-right(a-d) are the prediction result curves of the four quantum models on dataset S1, while from bottom-left to bottom-right(e-h) are that of quantum models on dataset S2}
\label{fig:curve}
\end{figure*}

\section{Background, challenge \& motivation}

\noindent\textbf{2.1 Background: Non-linearity in VQCs}
\vspace{5pt}

The strength of deep learning is its ability to solve complex problems by learning the complicated transformation between inputs and outputs through different neural network layers with non-linear functions. On the other hand, despite recent advances in quantum machine learning \cite{jiang2021co, liang2021can,wang2021exploration,wu2021scrambling,peddireddy2022classical, hu2022quantum,jing2022rgb,stein2022quclassi,hu2022design,gratsea2022exploring, zeng2022multi,wang2023qumos}, one of the major challenges is the ability to effectively handle nonlinearity. The model used in quantum machine learning itself is basically a linear transformation now given that the computation circuit doesn't involve any non-linearity but only matrix multiplication. However, many real-world phenomena cannot be adequately modeled using linear models. The research on non-linearity in QML becomes more important in order to enhance the ability of models. Therefore it can finally extend the frontier of this area in solving more difficult problems. Some existing works are making progress toward this goal, and one recent paper on the design of quantum circuits to help enable deep learning proposes a framework namely ST-VQC~\cite{li2023novel}. One important component of the framework is its data encoder that encodes the input data in groups, and then selectively duplicates the data groups to bring high-order polynomials as the duplicated data groups entangle, therefore introducing nonlinearity. Using its structure, we can see from Figure \ref{fig:curve} (a to d, from no duplicate to 3 duplicates) that duplicating the input has the potential to lead to better performance on the polynomial synthetic dataset S1 as we can see that the predicted curve(orange) can better fit the actual curve(blue).

\vspace{5pt}
\noindent\textbf{2.2 Challenge: No silver bullet exists while design space exploration for dedicate dataset has high cost}
\vspace{5pt}

As one example of the method proposed to bring nonlinearity to QML, despite its design's ability, ST-VQC, on the other hand, introduces a rather large exploration space for the hyperparameters. As a result, it comes with a cost in time regarding such a large search space. For example, there might be different choices among the number of duplications, the size of each group to be copied, and which group gets the most benefit from duplications. Although we can apply some techniques such as neural network search to explore the potential optimum hyperparameter settings, the time to execute the quantum circuits, on either simulation in classical computers or real quantum devices, counts for careful consideration. This situation could possibly get more serious as the number of involved qubits grows. For example, 100 epochs of normal training on the MNIST dataset can take up to 1.14 hours for a quantum neural network with 12 qubits and 82 parameters. For a more complicated quantum neural network with more qubits and parameters on larger datasets, it can cost days.

\begin{table}[]
\centering
\tabcolsep 4 pt
\small
\begin{tabular}{|cccccc|}
\hline
\multicolumn{1}{|c|}{Dataset} & \multicolumn{1}{c|}{Model} & \multicolumn{1}{c|}{\#Qubit} & \multicolumn{1}{c|}{\#Param} & \multicolumn{1}{c|}{RMSE} & \multicolumn{1}{c|}{R\textasciicircum{}2} \\ \hline
\multicolumn{1}{|c|}{S1} & \multicolumn{1}{c|}{ST-VQC(0 dup.)} & \multicolumn{1}{c|}{1} & \multicolumn{1}{c|}{48} & \multicolumn{1}{c|}{5.22E-01} & \multicolumn{1}{c|}{-1.125} \\ \cline{2-6} 
\multicolumn{1}{|c|}{} & \multicolumn{1}{c|}{ST-VQC(1 dup.)} & \multicolumn{1}{c|}{2} & \multicolumn{1}{c|}{48} & \multicolumn{1}{c|}{1.65E-01} & \multicolumn{1}{c|}{0.788} \\ \cline{2-6} 
\multicolumn{1}{|c|}{} & \multicolumn{1}{c|}{ST-VQC(2 dup.)} & \multicolumn{1}{c|}{3} & \multicolumn{1}{c|}{49} & \multicolumn{1}{c|}{1.68E-01} & \multicolumn{1}{c|}{0.780} \\ \cline{2-6} 
\multicolumn{1}{|c|}{} & \multicolumn{1}{c|}{ST-VQC(3 dup.)} & \multicolumn{1}{c|}{4} & \multicolumn{1}{c|}{48} & \multicolumn{1}{c|}{1.35E-01} & \multicolumn{1}{c|}{0.858} \\ \hline
\hline
\multicolumn{1}{|c|}{S2} & \multicolumn{1}{c|}{ST-VQC(0 dup.)} & \multicolumn{1}{c|}{1} & \multicolumn{1}{c|}{48} & \multicolumn{1}{c|}{6.41E-05} & \multicolumn{1}{c|}{0.999} \\ \cline{2-6} 
\multicolumn{1}{|c|}{} & \multicolumn{1}{c|}{ST-VQC(1 dup.)} & \multicolumn{1}{c|}{2} & \multicolumn{1}{c|}{48} & \multicolumn{1}{c|}{3.89E-03} & \multicolumn{1}{c|}{0.999} \\ \cline{2-6} 
\multicolumn{1}{|c|}{} & \multicolumn{1}{c|}{ST-VQC(2 dup.)} & \multicolumn{1}{c|}{3} & \multicolumn{1}{c|}{49} & \multicolumn{1}{c|}{3.24E-03} & \multicolumn{1}{c|}{0.999} \\ \cline{2-6} 
\multicolumn{1}{|c|}{} & \multicolumn{1}{c|}{ST-VQC(3 dup.)} & \multicolumn{1}{c|}{4} & \multicolumn{1}{c|}{48} & \multicolumn{1}{c|}{4.52E-03} & \multicolumn{1}{c|}{0.999} \\ \hline
\end{tabular}

\caption{Performance of different VQCs on synthetic datasets}
\label{tab:table1}
\end{table}

At the same time, it is better to take into consideration that there might be not always a universal solution to all datasets. As shown in Table \ref{tab:table1} and in Figure \ref{fig:curve}, the ST-VQC(3 dup.) works the best on S1, while all four models work similarly in performance on S2. On the other hand, it is also possible that one searched quantum circuit design performs well on one dataset while failing on another.

\vspace{5pt}
\noindent\textbf{2.3 Motivation: Benchmarking VQCs' approximability}
\vspace{5pt}


Motivated by this, there is a need for benchmarking quantum machine learning models on their ability to handle nonlinearity, which can contribute in several ways: 
\begin{itemize}[noitemsep,topsep=0pt,parsep=0pt,partopsep=0pt]
    \item Identify the strengths and weaknesses of different quantum machine learning models in dealing with nonlinearity. This information can guide the development of new models or the improvement of existing ones.
    \item Provide a quantitative measure of the performance of quantum machine learning models on nonlinear tasks, which can help evaluate and compare different models.    
    \item The benchmarking results can also provide insights into the types of problems that quantum machine learning models are well-suited for and those that may require further development. This can potentially guide the selection of appropriate quantum learning models for real-world applications.
\end{itemize}


\section{method}

\vspace{-15pt}

\begin{figure}[t]
    \centering
    \includegraphics[width=0.85\linewidth]{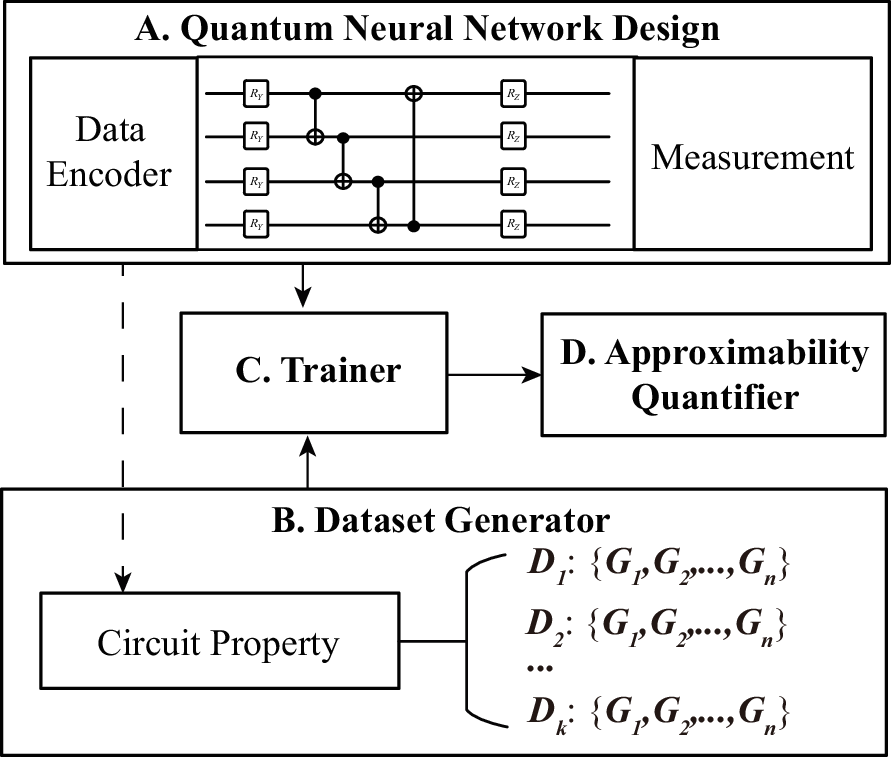}
    \caption{Overview of the proposed framework}
    \label{fig:overview}
\end{figure}

The proposed framework provides a comprehensive approach to benchmark the performance of quantum learning models. Figure \ref{fig:overview} illustrates an overview of our framework.
In the first step, the framework extracts the necessary property information from an input VQC design, which will then be utilized to generate a set of dataset groups. These datasets are designed to represent various degrees of nonlinearity by employing different polynomial functions. Next, the quantum neural network and the generated datasets are passed to the trainer to learn and adapt to the nonlinear relationships embedded in the data. To assess the model's performance in capturing nonlinearity, a metric called the "Approximability Score" is employed. This metric quantifies the model's ability to approximate the desired nonlinearity by collecting its performance on dataset groups.

\textbf{A. Quantum Neural Network Design.}
The input of the framework is VQC design from users. There are three main components in a VQC’s design: the data encoder, the computation circuit, and the measurement. Each component contributes to shaping the behavior and performance of the VQC model. The data encoder determines how the input data is transformed and encoded into the quantum states of the qubits. It encompasses various techniques such as amplitude encoding and angle encoding. The computation circuit defines the sequence of quantum gates and operations applied to the encoded data within the quantum model. It governs the flow of quantum information and interactions among the qubits. The measurement component determines how the final quantum state of the model is measured to obtain the output or prediction. When designing a VQC model within the framework, users have the flexibility to select and configure these components based on their specific requirements and the nature of the problem at hand. It allows customization and experimentation with different combinations of data encoders, computation circuits, and measurement techniques to optimize the model's performance.

\textbf{B. Dataset Generator.}
The dataset generation process will create datasets with different complexity and groups with different non-linearity for each dataset.
They will be used to estimate the non-linearity of the given VQCs.
By plotting a polynomial equation, we observe that the resulting curve can assume various shapes, which depend on the degree of the polynomial. For instance, a linear equation (degree 1) produces a straight line, while a quadratic equation (degree 2) generates a parabolic curve. As we increase the degree, more terms with higher exponents of the variable are introduced. This allows us to represent complex curves, including those with multiple peaks, valleys, or sharp turns, which are characteristic of highly nonlinear relationships.

To model the nonlinear relationship between the input and output in the datasets, we design them using polynomials with different degrees. Each dataset group is categorized based on the polynomial's degree, facilitating a systematic evaluation of the models' performance across varying degrees of nonlinearity. For example, a dataset group with a degree of 2 may consist of input variables, $x_0$ and $x_1$, and an output variable, $y$, defined as $y$ = $x_0^2$ + $x_1^2$. We create multiple datasets within each group, ensuring a diverse range of instances with distinct random coefficients for the polynomial terms.

This grouping and variation in polynomial degrees enable us to comprehensively assess the models' ability to capture and approximate nonlinear relationships across a spectrum of complexities. 
In addition, it is important to note that the input and output dimensions of the datasets are determined by the specific configuration of the data encoder and measurement components of the quantum model. For example, if the data encoder encodes the input variables into the states of qubits, the number of qubits used will define the dimensionality of the input space. Similarly, the choice of the measurement will influence the dimensionality of the output space.
By considering the structural characteristics of the quantum neural network design, we ensure that the generated datasets are compatible with the subsequent training and evaluation steps.

\textbf{C. Trainer.}
In the next step, we train the quantum neural network on each dataset within a specific group. The evaluation process involves executing the trained model on each dataset in the group, obtaining the Root Mean Squared Error ($RMSE$) and $R^2$ scores for each dataset. 

\textbf{D. Approximability Quantifier}
To quantify the overall performance of the VQC model on a dataset group, we define a metric that utilizes a weighted combination of the $RMSE$ and $R^2$ scores, assigning a weight of 0.5 to the (1-$RMSE$) term and a weight of 0.5 to the $R^2$ term. This weight reflects our emphasis on both the accuracy and the explained variance of the model's predictions.
The final score quantitatively measures the model's approximability for each dataset group, enabling a comparative analysis across different polynomial degrees and dataset complexities. A higher score indicates a better approximation of the nonlinearity, showcasing the model's capability to capture the underlying relationships.


\section{experimental results}
\begin{table}[]
\centering
\tabcolsep 5 pt
\small
\begin{tabular}{|c|c|c|cccc|}
\hline
 Dataset & Input & Output & \multicolumn{4}{c|}{NL Metric} \\ \hline
 &  &  & \multicolumn{1}{c|}{G1} & \multicolumn{1}{c|}{G2} & \multicolumn{1}{c|}{G3} & G4 \\ \hline
D1 & 2 & 1 & \multicolumn{1}{c|}{-0.258} & \multicolumn{1}{c|}{-0.329} & \multicolumn{1}{c|}{-0.416} & -0.497 \\ \hline
D2 & 4 & 1 & \multicolumn{1}{c|}{-0.38} & \multicolumn{1}{c|}{-0.391} & \multicolumn{1}{c|}{-0.471} & -0.997 \\ \hline
\end{tabular}
\caption{Dataset property}
\label{tab:dataset}
\end{table}

\noindent\textbf{4.1 Experimental Setup}\vspace{5pt}

Our experiments are conducted on 2 datasets: $D1$ and $D2$. In each dataset, there are 4 groups with different orders of non-linearity (i.e., polynomials): G1, G2, G3, and G4.
Note that the group index $X$ in $GX$ reflects the order of polynomials, which is $2X$; say G2 has 4 orders of polynomials.
In each group, there are 10 different combinations of polynomials, each of which contains 400 sample points.
We split these samples for training and testing, where the  ratio is 80\% for training and 20\% for testing.
The experiments on all datasets are trained with batch size 64 for 100 epochs, using the same optimizer and learning rate.


In order to better reflect the degree of nonlinearity existing in different dataset groups, we introduce an extra metric, the nonlinearity metric (NL Metric), as shown in Table \ref{tab:dataset}. 
The lower the NL Metric value, the more complex relationships exist in that dataset group.
For each dataset group, a linear model will be trained and tested along with the input quantum network to obtain its performance score. 
We can see that the linear model performs worse in the dataset groups with a high order of polynomials (i.e., larger group index).

\vspace{5pt}
\noindent\textbf{4.2 Benchmark Result}\vspace{5pt}

The benchmark results provide insights into the performance of the quantum machine learning models within the proposed framework. We conducted a comprehensive evaluation of various model designs. The goal was to assess the models' approximability, taking into account their performance on multiple dataset groups.
As shown in Table \ref{tab:result_table}, the first 4 columns represent the quantum circuits' properties- the number of qubits, parameters, the circuit depth, and the encoding method. The column "Dataset" indicates the dataset used for training, (e.g., D1 refers to Dataset 1 shown in Table \ref{tab:dataset}). Then the last four columns (G1 to G4) are the approximability scores on different degree groups.

We can extract useful information for analysis from such benchmark results. For example, in both datasets, we notice that the quantum models with amplitude encoding and angle encoding will suffer an accuracy drop when the dataset has higher orders of non-linearity. 
This is because the vanilla VQC only has 2-order polynomials, which are included during the measurement.
Then, according to \cite{li2023novel}, the ST-VQC can present higher-order polynomials by duplicating inputs.
If the number of duplications is 1 (denoted as ST-VQC(1 dup.)), it contains the 4-order polynomials, while duplication of 2 contains 8-order polynomials.
From the results, we can see the benchmark results are in accordance with the theoretic values.
On dataset D1, for ST-VQC(1 dup.), the best accuracy is on group G2, and it suffers an accuracy drop on G3 and G4; while ST-VQC(2 dup.) maintains the high accuracy on G4, which indicates it can represent the 8-order polynomials.
We also observe an exception on dataset D2, where ST-VQC(2 dup.) suffers an accuracy drop on G4. This may caused by the complexity of the dataset, and the number of parameters in ST-VQC(2 dup.) is not enough.
However, ST-VQC(2 dup.) still outperforms all other competitors on G4 in D2.
All these observations show the effectiveness of the proposed tool to estimate the approximability of a given VQC model.





\begin{table}[]
\centering
\tabcolsep 1 pt
\small
\label{tab:my-table}
\begin{tabular}{|ccccccccc|}
\hline
\multicolumn{1}{|c|}{\#Qub} & \multicolumn{1}{c|}{\#Para} & \multicolumn{1}{c|}{Dep.} & \multicolumn{1}{c|}{Enc.} & \multicolumn{1}{c|}{Dataset} & \multicolumn{1}{c|}{G1} & \multicolumn{1}{c|}{G2} & \multicolumn{1}{c|}{G3} & G4 \\ \hline
\multicolumn{1}{|c|}{1} & \multicolumn{1}{c|}{21} & \multicolumn{1}{c|}{21} & \multicolumn{1}{c|}{Amplitude} & \multicolumn{1}{c|}{D1} & \multicolumn{1}{c|}{0.999} & \multicolumn{1}{c|}{0.908} & \multicolumn{1}{c|}{0.724} & 0.509 \\ \hline
\multicolumn{1}{|c|}{2} & \multicolumn{1}{c|}{20} & \multicolumn{1}{c|}{15} & \multicolumn{1}{c|}{Angle} & \multicolumn{1}{c|}{D1} & \multicolumn{1}{c|}{0.954} & \multicolumn{1}{c|}{0.89} & \multicolumn{1}{c|}{0.745} & 0.549 \\ \hline
\multicolumn{1}{|c|}{2} & \multicolumn{1}{c|}{20} & \multicolumn{1}{c|}{15} & \multicolumn{1}{c|}{ST-VQC(1 dup.)} & \multicolumn{1}{c|}{D1} & \multicolumn{1}{c|}{0.993} & \multicolumn{1}{c|}{0.998} & \multicolumn{1}{c|}{0.976} & 0.938 \\ \hline
\multicolumn{1}{|c|}{3} & \multicolumn{1}{c|}{21} & \multicolumn{1}{c|}{15} & \multicolumn{1}{c|}{ST-VQC(2 dup.)} & \multicolumn{1}{c|}{D1} & \multicolumn{1}{c|}{0.966} & \multicolumn{1}{c|}{0.974} & \multicolumn{1}{c|}{0.984} & 0.979 \\ \hline
\multicolumn{1}{|l}{} & \multicolumn{1}{l}{} & \multicolumn{1}{l}{} & \multicolumn{1}{l}{} & \multicolumn{1}{l}{} & \multicolumn{1}{l}{} & \multicolumn{1}{l}{} & \multicolumn{1}{l}{} & \multicolumn{1}{l|}{} \\ \hline

\multicolumn{1}{|c|}{2} & \multicolumn{1}{c|}{20} & \multicolumn{1}{c|}{15} & \multicolumn{1}{c|}{Amplitude} & \multicolumn{1}{c|}{D2} & \multicolumn{1}{c|}{0.844} & \multicolumn{1}{c|}{0.707} & \multicolumn{1}{c|}{0.628} & 0.08 \\ \hline
\multicolumn{1}{|c|}{4} & \multicolumn{1}{c|}{32} & \multicolumn{1}{c|}{18} & \multicolumn{1}{c|}{Angle} & \multicolumn{1}{c|}{D2} & \multicolumn{1}{c|}{0.706} & \multicolumn{1}{c|}{0.669} & \multicolumn{1}{c|}{0.704} & -0.086 \\ \hline
\multicolumn{1}{|c|}{4} & \multicolumn{1}{c|}{32} & \multicolumn{1}{c|}{18} & \multicolumn{1}{c|}{ST-VQC(1 dup.)} & \multicolumn{1}{c|}{D2} & \multicolumn{1}{c|}{0.882} & \multicolumn{1}{c|}{0.879} & \multicolumn{1}{c|}{0.783} & 0.421 \\ \hline
\multicolumn{1}{|c|}{6} & \multicolumn{1}{c|}{42} & \multicolumn{1}{c|}{18} & \multicolumn{1}{c|}{ST-VQC(2 dup.)} & \multicolumn{1}{c|}{D2} & \multicolumn{1}{c|}{0.846} & \multicolumn{1}{c|}{0.808} & \multicolumn{1}{c|}{0.776} & 0.52 \\ \hline
\end{tabular}
\caption{Table for the benchmark result on different quantum machine learning model designs}
\label{tab:result_table}
\end{table}

\section{conclusion}
In this paper, we propose an automated tool to evaluate the ability of variational quantum circuits' approximability to different orders of polynomials, which reflects their ability to handle orders of nonlinearity.
The tool is composed of a dataset generator for a given VQC, a trainer, and an approximability quantifier.
With the proposed tool, the designers can efficiently explore the design space for a certain order of non-linearity without dedicating to a dataset.
What's more, it can be used for a fair comparison between the quantum version of learning models and the classical version with the same order of non-linearity.

\section*{Acknowledgment}
This work is partly supported by National Science Foundation (NSF) 2311949, Mason’s Office of Research Innovation and Economic Impact (ORIEI), and the U.S. Department of Energy, Office of Science, National Quantum Information Science Research Centers, Co-design Center for Quantum Advantage (C2QA) under contract number DE-SC0012704, (Basic Energy Sciences, PNNL FWP 76274).
The research used IBM Quantum resources via the Oak Ridge Leadership Computing Facility at the Oak Ridge National Lab, which is supported by the Office of Science of the U.S. Department of Energy under Contract No. DE-AC05-00OR22725, and IBM Quantum Hub at Los Alamos National Lab.





\bibliographystyle{IEEEbib}
\bibliography{refs}

\end{document}